\documentstyle[epsf]{mn}
\def\mch{M$\rm^{c}$Hardy\,}
\def\etal{\it et al.~\rm}

\begin{document}

\title{Simultaneous X-ray and infrared variability in the
quasar 3C273}

\author[I.M. M$\rm^{c}$Hardy \etal]
{Ian M$\rm^{c}$Hardy$^{1}$, Anthony Lawson$^{1}$, Andrew
Newsam$^{1,2}$, Alan Marscher$^{3}$, \and Ian Robson$^{4}$,
Jason Stevens$^{5}$
\\
$^{1}$ Department of Physics and Astronomy, University of Southampton,
SO17 1BJ.\\
$^{2}$ Liverpool John Moores University, Astrophysics Research Institute,
  Birkenhead CH41 1LD \\
$^{3}$ Department of Astronomy, Boston University, Boston, MA, 02215, USA.\\
$^{4}$ Joint Astronomy Center, Hilo, HI 96720, USA.\\
$^{5}$ Mullard Space Science Laboratory, UCL, Holmbury St Mary, Surrey,
RH5 6NT.}

\date{Accepted for publication in the MNRAS.}
\maketitle
\begin{abstract}
From a combination of high quality X-ray observations from the NASA
Rossi X-ray Timing Explorer (RXTE), and infrared observations from the
UK Infrared Telescope (UKIRT) we show that the medium energy X-ray (3-20
keV) and near infrared fluxes in the quasar 3C273 are highly
correlated. It is widely believed that the X-ray emission in quasars
like 3C273 arises from Compton scattering of low energy seed photons
and our observations provide the first reliable detection of
correlated variations in 3C273 between the X-ray band and any lower
energy band. For a realistic electron distribution we demonstrate that
it is probable that each decade of the seed photon distribution from
the mm to IR waveband contributes roughly equally to the medium energy
X-ray flux. However the expected mm variations are too small to be
detected above the noise, probably explaining the lack of success of
previous searches for a correlation between X-ray and mm variations.
In addition we show that the infrared leads the X-rays by
$0.75\pm0.25$ days.  These observations rule out the `External
Compton' emission process for the production of the X-rays.

\end{abstract}

\section{INTRODUCTION}

It is generally supposed that the high energy emission from blazars --
ie BL Lac objects and quasars which display some evidence of relativistic
jets -- arises from Compton scattering of low energy seed photons.
However the evidence for this supposition is quite weak.
There has been remarkably little progress, despite a great deal
of observational effort, in determining the details of the high energy
emission models. Various possibilities exist, all of which require
that the scattering particles are the relativistic electrons in the
jet.  The most popular hypothesis is the Synchrotron Self-Compton
(SSC) model in which the seed photons are the synchrotron photons from
the jet, up-scattered by their parent electrons. Alternatively the
seed photons may arise externally to the jet (the External Compton,
EC, process) or, in a combination of the two models, photons from the
jet may be mirrored back to the jet (the Mirror Compton, MC, model)
from a gas cloud before scattering up to high energies. The various
models make slightly different predictions about the lags between the
seed and Compton-scattered variations, and about the relative
amplitudes of the two components and so, in principle, the models can
be distiguished (eg see Ghisellini and Maraschi 1996 and Marscher 1996
for summaries of the predictions of the various models). Much
observational effort has therefore been devoted to attempting to find
correlated variability in the high and low energy bands.

\begin{figure*}
\begin{center}
  \leavevmode
  \epsfxsize 0.8\hsize
  \epsffile{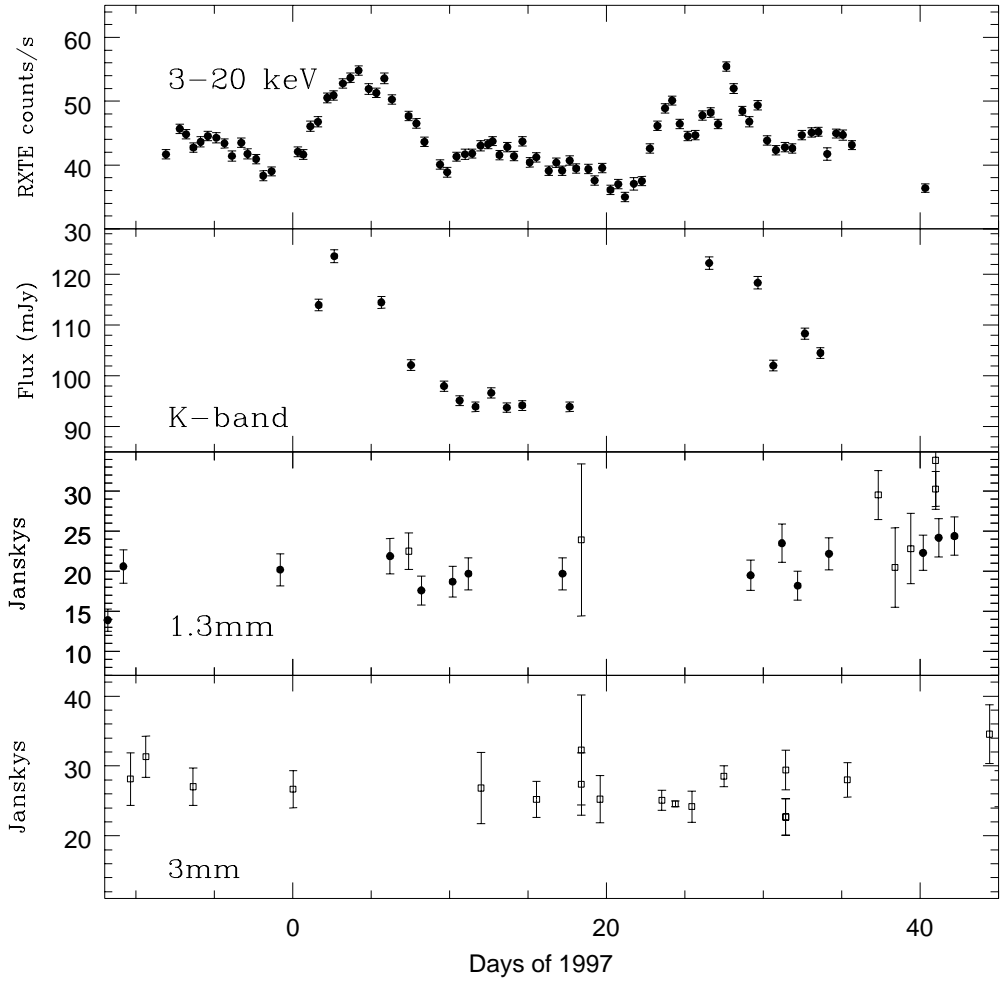} 
\end{center}
\caption{X-ray, infrared and millimetre lightcurves.
The X-ray counts are the total from 3 PCUs of the PCA.
The 1.3mm data are from the JCMT (filled circles), with some points from
OVRO (open squares).  The 3mm data are all from OVRO. }
\label{fig:lcurves}
\end{figure*}

In the SSC model, it has generally been expected that, as the peak of
the synchrotron photon number spectrum lies in the mm band for most
radio-selected blazars, the mm would provide the bulk of the seed
photons and so would be well correlated with the X-ray emission.
However in the case of 3C273, one of the brightest blazars, extensive
searches have been carried out for a connection between the X-ray and
millimetre bands on both daily (\mch 1993) and monthly (Courvoisier
\etal 1990; \mch 1996) timescales but no correlation has been
found. The SSC model may, however, be saved if the flaring synchrotron
component is self-absorbed at wavelengths longer than $\sim1$ mm. We
therefore undertook a search for a correlation between the X-ray 
and infrared emission in 3C273; previous observations (eg Courvoisier
\etal 1990; Robson \etal 1993) have confirmed that infrared flares in
3C273 are due to variations in a synchrotron component.  In the past,
large amplitude infrared flares have been seen only rarely in 3C273
(eg Courvoisier \etal 1990; Robson \etal 1993), partially because of
limited sampling which usually could not detect flares with overall
timescales $\sim$week. Nonetheless the previous sampling was
sufficient to show that such flare activity is not a continual
occurence.  It may be relevant that the present observations,
during which large amplitude infrared variability was detected,
were made during a period when the millimetre flux from 3C273 was very
high.

Here we present what we believe is the
best sampled observation of correlated variability between the
synchrotron and Compton-scattered wavebands in any blazar. The
observations cover not just one flaring event, which could be due to
chance, unrelated, flaring in the two wavebands, but two large
variations.  The observations, including the X-ray, infrared and
millimetre lightcurves, and cross-correlation of the X-ray and other
waveband lightcurves, are described in Section 2. The origin of the
X-ray seed photons is discussed in Section 3, the implications of the
observations are discussed in Section 4 and the overall
conclusions are given in Section 5.

\section{OBSERVATIONS} 

\subsection{X-ray Observations}

During the 6 week period from 22 December 1996 to 5 February 1997,
X-ray observations were carried out twice a day by RXTE and nightly near
infrared service observations were made at the United Kingdom Infrared
Telescope (UKIRT).

The X-ray observations were made with the large area (0.7 m$^{2}$)
Proportional Counter Array (PCA) on RXTE (Bradt, Rothschild
and Swank 1993). Each observation lasted for
$\sim1$ksec.  The PCA is a non-imaging device with a field of view of
FWHM $\sim1^\circ$ and so the background count rate was calculated
using the RXTE q6 background model. Standard selection
criteria were applied to reject data of particularly high background
contamination. 

3C273 is detectable in each observation in the energy range 3-20 keV
and its spectrum is well fitted by a simple power law.  As with other
PCA spectra (eg The Crab --see
http://lheawww.gsfc.nasa.gov/users/keith/pcarmf.html) the measured
energy index, $\alpha$=0.7, is 0.1 steeper than measured by previous
experiments, eg GINGA (Turner \etal 1990).  The X-ray spectra, and
spectral variability during the present observations are discussed in
detail by Lawson \etal (in preparation).  The average count rate of 45 counts
s$^{-1}$ (3-20 keV) (the total for 3 of the proportional counter units,
PCUs, of the PCA) corresponds to a flux of $1.5 \times 10^{-10}$ ergs
cm$^{-2}$ s$^{-1}$ (2-10 keV).

In figure~\ref{fig:lcurves} we present the count rate in the 3-20 keV
band. We see two large X-ray flares. The first flare begins on
approximately 1 January 1997, reaches a peak on 4 January and returns
to its pre-flare level on 10 January. The flare is quite smooth. The
second flare begins on 22 January and lasts until approximately 1
February. The initial rise is faster than that of the first flare, and
the overall shape indicates a superposition of a number of smaller
flares. X-ray spectral variations are seen during the flares (Lawson
\etal in preparation), showing that changes in the Doppler factor of the jet
cannot, alone, explain the observed variability.

\subsection{Infrared and Millimetre Observations}

In figure~\ref{fig:lcurves} we show 1.3 and 3 mm observations from the
James Clerk Maxwell Telescope (JCMT - see Robson \etal 1993 for
reduction details) and from the Owens Valley Radio Observatory
(OVRO); the latter data were obtained from the calibration
database.  There is no evidence of flares of comparable amplitude to
those in the X-ray lightcurve, but the sampling is poorer and the
errors are larger.

We also show the K-band lightcurve derived from service observations
at the United Kingdom Infrared Telescope (UKIRT) from 1 January until
3 February 1997. The observations were made with the infrared imaging
camera IRCAM3 with typical exposures of 3 minutes. The observations
were made in a standard mosaic manner and the data were also reduced
in a standard manner. There are some gaps due to poor weather but
increases in the infrared flux at the same time as the X-ray flares
can be seen clearly.  The average K error is $\sim1$mJy (ie 1 per
cent). Approximately half of the error comes from the Poisson noise
and the rest comes from calibration uncertainties.

\subsection{X-ray/Infrared Cross-Correlation}

\begin{figure}
\begin{center}
  \leavevmode
  \epsfxsize 1.0\hsize
  \epsffile{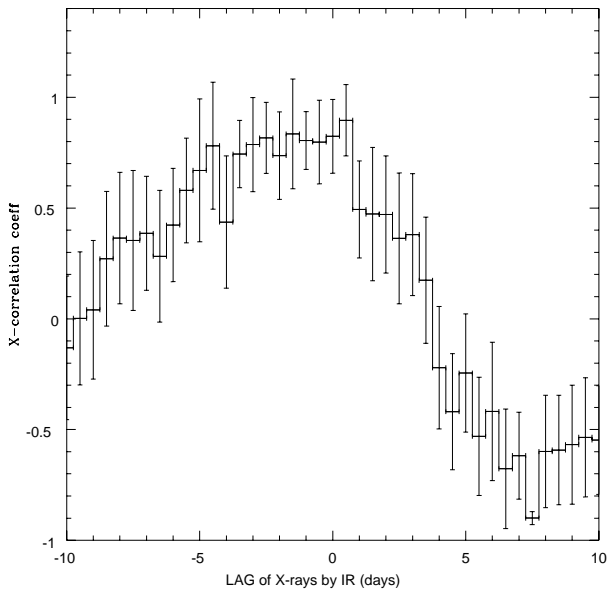} 
\end{center}
\caption{Cross-correlation of the 3-20 keV X-ray lightcurve and
K-band lightcurves shown in figure~\ref{fig:lcurves}.
}
\label{fig:xcor}
\end{figure}

We have cross-correlated the X-ray lightcurves with the millimetre and
K-band lightcurves using the Edelson and Krolik (1988) discrete
cross-correlation algorithm as coded by Bruce Peterson (private communication).
As found previously there is no correlation of the
X-ray emission with the millimetre emission but there is a very strong
correlation with the infrared emission (figure~\ref{fig:xcor}) with
correlation coefficient close to unity. The cross-correlation peaks
close to zero days lag but is asymetric. Although we can rule out the
infrared lagging the X-rays by more than about one day, a lag of the
infrared by the X-rays by up to 5 days is possible.

The observations presented here are the first to show a definite
correlation in 3C273 between the X-ray emission and that of any 
potential seed photons.

\section{THE ORIGIN OF THE X-RAY SEED PHOTONS}

An important question is whether the infrared photons are actually the
seed photons for the X-ray emission or whether they are simply tracers
of a more extended spectral continuum, with the X-rays arising from
scattering of another part of the continuum. Robson \etal (1993) state
that in 3C273 the onset and peak of flares occur more or less
simultaneously (ie lags of $<1$ day) from K-band to 1.1 mm.
Therefore although we have not adequately monitored at wavelengths
longer than 2.2$\mu$, we assume that the whole IR to mm continuum does
rise simultaneously. 

\begin{figure}
\begin{center}
  \leavevmode
  \epsfxsize 1.0\hsize
  \epsffile{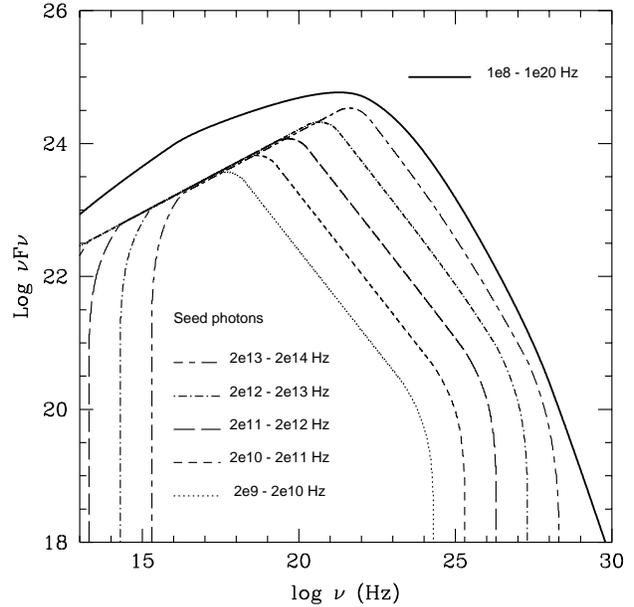} 
\end{center}
\caption{Compton scattered spectrum resulting from the scattering of
a seed photon spectrum stretching from $\nu=10^{8}$ to $10^{20}$ Hz.
At low frequencies the spectral index, $\alpha$, where flux,
$F, \propto \nu^{-\alpha}$, is 0.75 and, above a break frequency
of $10^{14}$ Hz, $\alpha = 1.5$. The electron number energy spectrum,
$N(\gamma) \propto \gamma^{-m}$, where $\gamma$ is the Lorentz factor
of the individual electrons, stretches from $\gamma=10$ to $10^{7}$,
with a slope, $m$, at low energies of 2.5 and $m=4.0$ above
$\gamma=10^{4}$. The proper Klein-Nishina cross section is used.
No bulk relativistic motion is included.
The thick line represents the total scattered spectrum. The
other lines represent the result of scattering seed photons with
only one decade of energy. 
Note that, in the medium energy X-ray band (4 keV= $10^{18}$ Hz), seed
photons from all decades from cm to near infrared contribute equally
to the scattered flux, with each contributing about 20 per cent.}
\label{fig:scatter}
\end{figure}

We have therefore calculated the Compton scattered spectrum resulting
from the scattering of individual decades of seed photon energies,
from the infrared to millimetre bands. The seed photons are taken from
a typical photon distribution and are scattered by a typical electron
distribution.  The resulting scattered spectra are shown in
figure~\ref{fig:scatter} and details of the photon and electron
distributions are given in the caption to figure~\ref{fig:scatter}.
It is assumed that the emission region is optically thin which, in
blazars, is true for the large majority of frequencies discussed in
figure~\ref{fig:scatter}.  Note that although the electron and input
photon spectra are self-consistent as regards the SSC mechanism, the
result is general and applies to scattering of seed photons produced
by any mechanism.  At the highest Compton scattered energies, ie GeV,
only the highest energy seed photons below the break in the photon
distribution (ie near infrared) are important. However at medium
energy X-rays we get approximately equal contributions from each
decade of seed photons. Thus scattered infrared photons probably
contribute about 20 per cent of the medium energy X-ray flux and the sum of
the scattered X-ray emission from lower energy seed photons exceeds
that from the infrared alone. These ratios can be altered slightly by
different choices of seed photon and electron spectral index, but the
general result is robust.

If the infrared is indeed a tracer of the seed photon continuum, we
can extrapolate to find the expected variability in the millimetre
band.  The peak and minimum observed K fluxes during our observations
are 124 and 93 mJy respectively, ie a range of 31 mJy, although we
note that we do not have K observations at either the peak or minimum
of the X-ray lightcurves and so the true range of K-band variability
may be somewhat more.  If the spectral index, $\alpha$, of the seed
spectrum is 0.75 (as reported by Robson \etal and Stevens \etal 1998)
we would then expect a rise of $\sim$3.7 Jy at 1.3 mm, which we cannot
rule out in the present observations and which would not have been
easy to detect in previous, less well sampled, monitoring
observations, explaining the lack of success of previous searches for
millimetre/X-ray correlations.  At 3mm the predicted variability
amplitude would be 7 Jy.  Robson \etal states that the 3mm rises lag
1mm rises by about 6 days, and 3mm decays are substantially longer,
which would all make them easier to detect, given our sampling
pattern.  However, with the exception of the very last datapoint at
day 44, no deviations of more than 5 Jy from the mean level are
detected. The implication is that $\alpha \leq 0.75$ or that the
flaring component is self absorbed by 3mm.  If the flaring component
has $\alpha=1.2$ as derived for the 1983 flare by Marscher and Gear
(1985), that component would have to be self absorbed by 1.3mm.

\section{DISCUSSION}

There are two major observational constraints on the X-ray emission 
mechanism: the relative amplitudes of the synchrotron and Compton
scattered components, and the time lag between them. Here we 
attempt to constrain these parameters by modelling the X-ray lightcurve.

\subsection{Modelling the X-ray lightcurve}

If the X-ray emission is physically related to the infrared emission,
then we can parameterise the relationship by:

\[ X_{predicted}(t)= A \, (K_{flux}(t-\delta t) - K_{quiescent})^{N} 
\, + \, X_{quiescent} \]

$K_{quiescent}$ is a non-varying K-band component. Robson \etal (1993)
show that such a component, steady on a timescale of years, is
contributed probably by warm dust in the broad line clouds,
heated to the point of evaporation.
Following Robson \etal we fix $K_{quiescent}=50$mJy.
$K_{flux}(t-\delta t)$ is the total observed K-band flux at time
$t-\delta t$ and $X_{predicted}(t)$ is then the predicted total X-ray
flux at time $t$.  $X_{quiescent}$ is the part of the X-ray flux which
does not come from the flaring region. The variable $\delta t$ is
included to allow for lags between the X-ray and infrared variations.
Initially we set $\delta t = 0$ but, in section 4.2, we consider the
implications of allowing $\delta t$ to vary.
$A$ is the constant of proportionality
(containing information about the electron
density, magnetic field and the various flux conversion constants)
and $N$ contains information about the emission mechanism.  For
example if the X-rays arise from variations in electron density then
we expect $N=2$ in the SSC and MC processes, but in the EC model $N=1$.
We have therefore performed a $\chi^{2}$ fit, using a standard
Levenburg-Marquardt minimisation routine, comparing the predicted
X-ray flux with the observed flux, in order to determine the three
unknowns, $A$, $X_{quiescent}$ and $N$. The errors on the predicted
X-ray flux are derived from the observed errors on the infrared flux.

The present infrared lightcurve is not well enough sampled to
determine all 3 parameters independently but, if $X_{quiescent}$ could
be determined precisely from other observations, then we could
determine $N$ to $\pm0.2$. Here $N$ varies from 0.5 for
$X_{quiescent}=0$ to 1.0 for $X_{quiescent}=23$ and 2.0 for
$X_{quiescent}=35$.  The minimum observed value of the total X-ray
count rate during the present observations was 35 count s$^{-1}$.
Hence as some part of those 35 count s$^{-1}$ almost certainly comes
from X-ray components which are not associated with the flaring
activity, eg a Seyfert-like nucleus or other parts of the jet, then
the maximum allowed value of $N$ is probably just below 2.  Typical
RXTE count rates outside of major flaring periods are in the range
20-25 counts s$^{-1}$ and fluxes observed by previous satellites (eg
see Turner \etal 1990) correspond to the same flux range. If that
count rate represents the true value of $X_{quiescent}$, then $N$ is
probably nearer unity, favouring EC models, or SSC or MC models in
which variations in the magnetic field strength play an important part
in flux variations.

\subsection{Implications of lightcurve modelling for lags}

\begin{figure}
\begin{center}
  \leavevmode
  \epsfxsize 1.0\hsize
  \epsffile{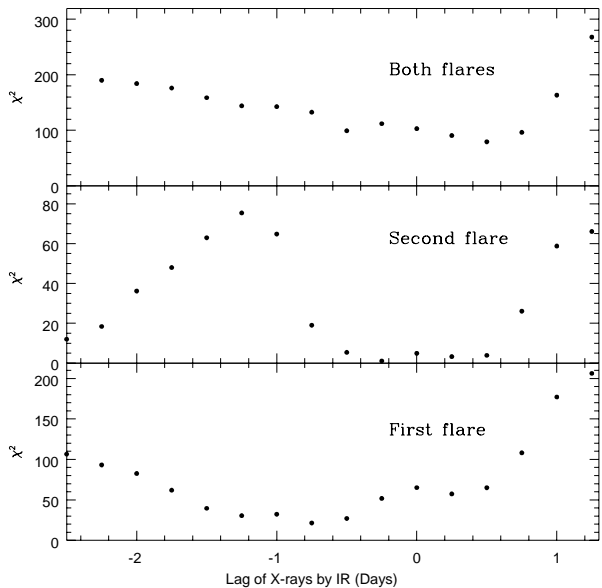}  
\end{center}
\caption{
Results of comparing the observed X-ray lightcurve with that predicted
from the infrared variations, with all parameters allowed to remain
free apart from the X-ray/infrared lag. The numbers of degrees of 
freedom are 13 (both flares), 8 (first flare) and 2 (second flare).
Note that it is impossible
to obtain a good fit to both X-ray flares simultaneously but acceptable
fits can be obtained to both fits individually. However the lags are
different for the two flares with the X-rays lagging the infrared by
$\sim0.75$ days in the first flare but the X-rays and infrared being
approximately simultaneous in the second flare.
}
\label{fig:lagschi}
\end{figure}

Comparison of the best-fit predicted and observed X-ray fluxes reveals
that, in the first flare, the predicted fluxes exceed the observed
fluxes on the rise and the reverse is true on the fall.  A better fit,
at least for the first flare, occurs if the predicted lightcurve is
delayed by about a day (in other words, the observed IR leads the
X-rays).  We therefore introduced a variety of time shifts, $\delta
t$ above,
into the IR lightcurve, and also separately considered the first
and second flares, and refitted. We applied simple linear
interpolation to estimate the IR flux at the exact (shifted) time of
the X-ray observations.  The results are shown in
figure~\ref{fig:lagschi}.

When considering all of the IR data, we obtain a plot (top panel of
figure~\ref{fig:lagschi}) which is rather similar to the
cross-correlation plot (figure~\ref{fig:xcor}), which is not too
surprising as the analysis techniques are similar, although the
modelling in principle allows us to quantify the goodness of fit.
We are cautious of overinterpreting the above datasets and so, we
prefer to plot figure~\ref{fig:lagschi} in terms of raw $\chi^{2}$
rather than probabilities which might be taken too literally.  As in
many analyses where the errors are small, slight (real) differences in
data streams lead to low probabilities of agreement even though
overall agreement is very good. Here a minor variation in either
X-rays or IR from a region not associated with the flare could
provide that small difference. However the change in relative
goodness of fit can be easily seen from the $\chi^{2}$ plots.
When we consider separately the IR data from the first flare
(ie the 11 data points up to day 20 of 1997), or from the second flare
(the remaining 5 data points) we obtain much better fits. We find
that the first flare is best fitted if the IR leads the X-rays by
about 0.75 days.  We are again cautious in ascribing exact errors
to the lag but changes of $\delta \chi^{2}$ of 6.4, corresponding
to 40 per cent confidence, occur in the first flare at 0.25 days from
the minimum value. A lag of the X-rays by the IR by less
than 0.25 days is ruled out at the 99.97 per cent confidence level.
The more limited data of the second flare is, however,
best fitted by simultaneous IR and X-ray variations.

Again with caution, we note that Lawson \etal (in preparation) find
different X-ray spectral behaviour between the two flares. In the
first flare the spectrum hardens at the flare onset but, at the peak,
the spectrum is softer than the time averaged spectrum; in the second
flare the hardness tracks the flux quite closely with the hardest
emission corresponding to the peak flux. Thus there do appear to be
differences between the two flares. However whether the observed
differences are due to differences in, for example the physical
parameters of the emitting region (eg density, magnetic field
strength), the strength of any exciting shock, or the geometry of the
emitting regions, is not yet clear but is an interesting subject
for future investigations.

Although not really intended for such analysis, blind application of
Keith Horne's Maximum Entropy Echo Mapping software to the whole
dataset also leads to the conclusion that the IR leads the X-rays by
0.75 days (Horne, private communication).

\begin{figure}
\begin{center}
  \leavevmode
  \epsfxsize 1.0\hsize
  \epsffile{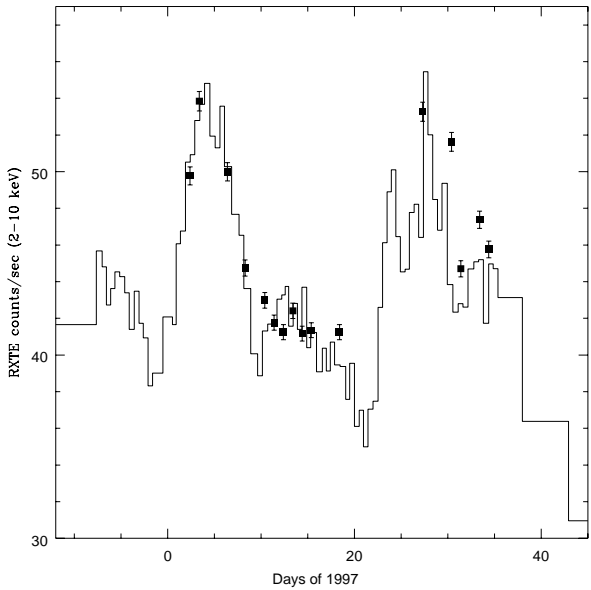} 
\end{center}
\caption{Observed X-ray lightcurve (histogram) and the best fit
predicted X-ray flux (filled squares) based on the parameters derived
from fitting the infrared observations to the first flare (11 data points)
only. The best-fit parameters are $A=0.47$, $N=0.98$ and $X_{quiescent}=22.1$.
Following from figure~\ref{fig:lagschi} the lead of the
IR over the observed X-rays is fixed at 0.75 days,
the best-fit value for the first flare.
The observed X-ray errorbars (see figure~\ref{fig:lcurves})
are not repeated here to avoid cluttering the diagram.
Note how the predicted X-ray fluxes for the second `flare' are then
systematically overestimated and also slightly lag the observed 
X-ray fluxes.}
\label{fig:xpred}
\end{figure}

As an example we show, in figure~\ref{fig:xpred}, the observed X-ray
lightcurve and the predicted lightcurve, based on parameters derived
from fitting to just the first flare with the IR leading by 0.75 days
(the best fit). We see that such a lag does not fit the second flare
well. In particular the predicted X-ray fluxes for the second flare
all lie above the observed fluxes by about 4 counts s$^{-1}$ and the
predicted fluxes now slightly lag (by about half a day) the observed
fluxes. One possible explanation of the excess is that $X_{quiescent}$
is lower during the second flare.  From our long term weekly
monitoring (in preparation) we note that the two flares shown here are
actually superposed on a slowly decreasing trend of the correct slope
to explain the excess. Inclusion of such a trend into our fitting
procedure does produce a slightly better fit for the overall dataset,
but the different lags between the first and second flare still
prevent a good overall fit from being obtained.  We therefore favour
the explanation that the long term lightcurve is actually made
up of a number of short timescale (week) flares, superposed on a more
slowly varying (months) `quiescent' component, rather than proposing
that the lightcurve is made up entirely of short flares, with no
underlying `quiescent' component.

\subsection{The X-ray Emission Mechanism}

The similarity between the present infrared variations and those of
previous infrared variations where the whole IR to mm continuum varied
together (Robson \etal 1993), and the lack of any other likely source
of rapidly variable infrared radiation, means that the varying
component of the infrared flux is almost certainly synchrotron
radiation from the jet. The very strong correlation between the X-ray
and infrared lightcurves shows that the same electrons which produce
the infrared synchrotron emission must also produce the scattered
X-ray emission.  The original version of the EC
model (Dermer and Schlickeiser 1993)
in which the high energy variations are
caused by variations in the external seed photons is thus ruled out.
The next version, in which the electrons in the jet which produce the
infrared synchrotron emission also scatter an all-pervading ambient
nuclear photon field (Sikora, Begelman and Rees 1994)
is also ruled out, at least for the first flare,
as we would then expect exactly simultaneous  X-ray and
infrared variations.

The remaining possible emission mechanisms are the SSC process,
which must occur at some level, and the MC process. In the SSC process
we expect, for moderate variations such as those observed here where
the emission region probably remains optically thin, that the X-ray
flares will lag the IR flares (in the source frame) by approximately
the light travel time across the radius of the emission region. The lag is 
because most photons will not be scattered where they
are produced but will typically travel the radius of the emission
region before being scattered.  In this model we can therefore deduce
the radius if we know the bulk Lorentz factor of the jet.
In the MC model the low energy photons also lead the high energy
photons, in this case by approximately the light travel time between
the emission region in the jet and the cloud.
If the cloud forms part of the broad line region we
might reasonably expect lags of order days. 

The EC model is ruled out by the IR/X-ray lag but both the SSC and MC
models are consistent with the lag. The parameter $N$ is not yet well
defined but the present indications are that it is closer to 1 than to
2, which, for the SSC and MC models, implies that changes in magnetic
field strength are at least partially responsible for the observed
variations.  The MC compton scattered flux has a higher dependence on
the bulk Lorentz factor of the jet than does the SSC mechanism, but
that factor is very hard to measure.

\section{CONCLUSIONS}

We have demonstrated, for the first time, a strong relationship between
the X-ray emission and that in any other lower frequency band in
3C273.  We have shown that the IR and X-ray emission in 3C273 are very
strongly correlated. By means of a simple calculation we have shown that
each decade of the synchrotron spectrum from the cm to IR bands probably
contributes equally (at about 20 per cent per decade) to the Compton scattered
X-ray flux. Overall the lag between the IR and X-ray bands is very small
but, in at least the first flare, the IR 
leads the X-ray emission by $\sim0.75\pm0.25$ days.
This lag rules out the EC model but is consistent with either the
SSC or MC model.

We have attempted to measure the parameter $N$ which determines the
relationship between the seed photon and Compton
scattered flux. The present data do not greatly constrain $N$ although
they indicate that 2 is the absolute upper limit and that a lower
value is probable. In terms of the SSC or MC models the implication is
that changes in the magnetic field strength are responsible for
at least part of the observed variations and, for  $N=1$, could
be responsible for all of the variations.

Because of their intrinsic similarity, the SSC and MC models
are hard to distinguish. However if it were possible to measure
IR/X-ray lags for a number of flares, of similar amplitude, in the
same source, then in the SSC model one would expect broadly similar
lags in each case, assuming that the emission comes from physically
similar emission regions. However in the MC model the reflecting
clouds will probably be at a variety of different distances and
so the lags should be different in each case. 

We may also examine variations in optical and UV emission line
strength. If synchrotron radiation from the jet is irradiating
surrounding clouds (MC process), then we would expect the resultant
recombination line radiation to vary with similar amplitude to, and
simultaneously with, the synchrotron emission. However in
the SSC process we would expect no change in emission line
strength.

Further X-ray/IR observations with $\sim$few hour time resolution
are required to refine the lag found here and to determine whether
the lag is different in different flares.
\\

{\bf Acknowledgements} We are very pleased to thank the management and
operatonal staff of both RXTE and UKIRT for their cooperation in
scheduling and carrying out these observations. 
We thank Keith Horne for running our data through his MEMECHO software. 
IM$\rm ^{c}$H thanks PPARC for grant support
and APM was 
supported in part by NASA Astrophysical Theory Grant NAG5-3839.

{}


\begin{thebibliography}{}

\bibitem[]{}Bradt, H.V.D, Rothschild, R.E. and Swank, J.H., 1993.
A\&AS, 97, 355.

\bibitem[]{}Courvoisier, T.J-L. \etal 1990. A\&A 234 73.

\bibitem[]{}Dermer, C.D. and Schlickeiser, R., 1993. ApJ, 416, 458.

\bibitem[]{}Edelson, R.A. and Krolik, J., 1988. ApJ, 333, 646.

\bibitem[]{}Ghisellini, G. and Maraschi, L., 1996. ASP Conf. Series, 110, 436.

\bibitem[]{}Marscher, A.P. and Gear, W.K., 1985. ApJ, 298, 1114.

\bibitem[]{}Marscher, A.P., 1996. ASP Conf. Series, 110, 248. 

\bibitem[]{}M$\rm^{c}$Hardy, I.M., 1993. Proc IAU Symposium 159, 
eds Courvoisier, T.  and Blecha, A., Kluwer, p193.

\bibitem[]{}M$\rm^{c}$Hardy, I.M., 1996. ASP Conf. Series, 110, 293.

%\bibitem[]{}Robson, I., Gear, W., Brown, L., Courvoisier, T., 
%Smith, M., Griffin, M. and Blecha, A., 1986. Nature 323, 134.

\bibitem[]{}Robson, E.I. \etal, 1993. MNRAS, 262, 249.

\bibitem[]{}Sikora, M. and Begelman, M.C. and Rees, M.J., 1994. ApJ,
421, 153.

\bibitem[]{}Stevens, J., Robson, I., Gear, W., Cawthorne, T., 
Aller, M., Aller, H., Terasranta, H. and Wright, M., 1998, ApJ
502, 182.

\bibitem[]{}Turner, M.J.L. \etal 1990. MNRAS, 

%\bibitem[]{}Unwin, S., Wehrle, A., Lobanov, A.P., Zensus, J.A., 
%Madejski, G.M., Aller, M.F. and Aller, H.D., 1996. ApJ 480, 596.

\end{thebibliography}
\end{document}